\documentclass[a4paper, prb, twocolumn, amsmath, amssymb, floatfix,superscriptaddress]{revtex4}
\usepackage{graphicx}
\usepackage{nicefrac}
\usepackage{epstopdf}
\graphicspath{{./pdfs/}}
\usepackage{dcolumn}					
\usepackage{bm}							
\usepackage{color}
\usepackage[pdfauthor={Peter Kratzer},pdftitle={title},pdfsubject={subject}, colorlinks=false, citecolor=black, linkcolor=black, bookmarksopen]{hyperref}

\begin{document}
\def\be{\begin{equation}}
\def\ba{\begin{array}}
\def\bea{\begin{eqnarray}}		
\def\ee{\end{equation}}
\def\ea{\end{array}}
\def\eea{\end{eqnarray}}		
\def\nn{\nonumber \\}
\def\inte{\int \limits}
\preprint{relaxation}

\title{Relaxation of photoexcited hot carriers beyond multi-temperature models: General theory description verified by experiments on Pb/Si(111)}

\author{Peter Kratzer}
\affiliation{Fakult{\"a}t f{\"u}r Physik and Center for Nanointegration (CENIDE), Universit{\"a}t Duisburg-Essen, D-47048 Duisburg, Germany}

\author{Laurenz Rettig}
\affiliation{Fritz-Haber-Institut der Max-Planck-Gesellschaft, D-14195 Berlin, Germany}

\author{Irina Yu. Sklyadneva}
\affiliation{Donostia International Physics Center (DIPC), 20018 San Sebastian/Donostia, Basque Country, Spain}

\author{Evgueni V. Chulkov}
\affiliation{Donostia International Physics Center (DIPC), 20018 San Sebastian/Donostia, Basque Country,  Spain}
\affiliation{Departamento de Polimeros y Materiales Avanzados: Fisica, Quimica y Tecnologia, Facultad de Ciencias Quimicas,
Universidad del Pais Vasco UPV/EHU, 20080 San Sebastian/Donostia, Spain }

\author{Uwe Bovensiepen}
\affiliation{Fakult{\"a}t f{\"u}r Physik and Center for Nanointegration (CENIDE), Universit{\"a}t Duisburg-Essen, D-47048 Duisburg, Germany}


\begin{abstract}
The equilibration of electronic carriers in metals after excitation by an ultra-short laser pulse provides an important 
class of non-equilibrium phenomena in metals and allows measuring the effective electron-phonon coupling parameter. Since the observed decay of the electronic distribution is governed by the interplay of both electron-electron and electron-phonon scattering, the interpretation of experimental data must rely on models that ideally should be easy to handle,  yet accurate. 
In this work, an extended rate-equation model is proposed that explicitly includes non-thermal electronic carriers while at the same time 
incorporating data from first-principles calculations of the electron-phonon coupling via Eliashberg-Migdal theory.  The model is verified against experimental data for thin Pb films grown on Si(111). Improved agreement between theory and experiment at short times ($<0.3$ps) due to non-thermal electron contributions is found. Moreover, the rate equations allow for widely different coupling strength to different phonon subsystems. Consequently, an indirect, electron-mediated energy transfer between strongly and weakly coupled groups of phonons can be observed in the simulations that leads to a retarded equilibration of the subsystems only after several picoseconds.  
\end{abstract}

\maketitle

\section{Introduction}

Ultra-fast spectroscopies, especially pump-probe experiments, are one of the most powerful techniques to explore the non-equilibrium behavior of charge carriers in condensed matter. Already very early in the development of this method, the time-dependent distribution of hot electrons and holes in metals has come into focus.\cite{FaSt92,GrSp95} 
For the interpretation of the time-dependent spectra, the 
two-temperature model \cite{AnKa74} has become an invaluable tool. 
It is employed to describe the equilibration between the electronic excitations and the lattice degrees of freedom, using thermal distribution functions for both. 
Recent extensions to multi-temperature models in which the lattice vibrations are described by multiple heat baths at various temperatures~\cite{Waldecker16,MaCa17} can account for the largely varying coupling strengths between electrons and vibrations of different polarization and/or frequency.
Yet, shortcomings of such models have been noticed \cite{GrSp95} already in the early days of the field:
Building on the concept of a time-dependent electronic temperature implies that this temperature drops in a linear fashion (if the electronic density of states is assumed constant), which is often not in accord with experimental observations.\cite{LiLo04,ObDe20} 
Moreover, the time required for the electronic temperature to reach the equilibrium temperature should, according to models of this type, be proportional to the amount of energy deposited. Again, this is not what is observed experimentally. 

All models of this type rest on the common assumption that 
electron-electron (e-e) scattering is much more efficient (at least for carriers far away the Fermi energy) than electron-phonon (e-ph) scattering. 
Due to this, the electrons and holes would first establish equilibrium among themselves\cite{ReKa02} (within less than 1 ps after excitation),
thus allowing us to start from an 'gas' of carriers at a well-defined, but time-dependent temperature $T_e(t)$ to initiate the flow of energy from the electronic system to the lattice. 
Meanwhile, we are well aware\cite{MuRe14} that this is an oversimplification; various more sophisticated theoretical approaches have indicated 
overlapping time scales of e-e and e-ph scattering. For instance, this can be demonstrated in the context of a Hubbard model coupled to lattice vibrations by applying the formalism of non-equilibrium Green functions \cite{RaFr16,KemperFreericks2018}. 
Moreover, calculations of the self-energy related to electron-phonon interaction \cite{CaNo20,CaNo22} or microscopically derived rate-equation models \cite{BaKa14,KrZa19}  can be used to 
incorporate data from first-principles calculations. The results of such simulations indicate that overlapping time scales are relevant for realistic systems. 

In this work, we build a model for relaxation in a metallic film and present the results of simulations that are compared to experimental data. 
Similar to multi-temperature models, our rate-equation model is (in most of its parts) analytical and easy to use, yet capable of describing the deviations from thermal distributions on {\em both} the electronic and the lattice degrees of freedom. 
For this purpose, we use a simple but flexible starting point to describe the electronic system, consisting of  an excess contribution $h_k(t)$ of hot carriers on top of a Fermi distribution function at a temperature $T_e(t)$, 
\be
f^{(0)}_k(T_e(t)) = \left[ \exp\left( \frac{\varepsilon_k - \mu}{k_B T_e(t)} \right) +1 \right]^{-1}
\ee
Here $k$ stands for all quantum numbers characterizing an electronic state and $\mu \approx E_F$ is the chemical potential of the charge carriers. After their equilibration with the lattice, a common temperature is reached, $T_e(t) \to T_\infty$. 
The objective of the proposed generalization is to account for possibly overlapping time scales of e-e and e-ph scattering found in metals. 
In metals or semimetals, the impact ionization and Auger recombination processes \cite{Tomadin2013}, involving electronic states both below and above the Fermi energy, make the dominant contribution to Coulomb scattering. 
This is different from (wide-gap) semiconductors in which the Coulomb interaction among photoexcited carriers leads to intra-band scattering among the electrons in the conduction band, or among the holes in the valence band, but scattering involving {\em both}  electrons and holes is inefficient due to the energetic separation between them. 
In a metal,  an expansion of the electronic distribution into eigenfunctions of the scattering kernel originating from the screened Coulomb interaction can be employed. 
Kabanov and Alexandrov \cite{KaAl08} 
started from the Boltzmann equation and showed that the Fermi distribution is just one possible solution describing a quasi-equilibrium state, but fast temporal changes require higher eigenfunctions of the scattering kernel to be included. 
This leads us to propose the ansatz 
\bea
f_k(t) &=& f^{(0)}_k(T_e(t)) + h_k(t) \, ,
\label{eq:hk} \\
h_k(t) & \equiv & 
h(\varepsilon_k,t) = h_0 \, \xi \, \exp\left( - \frac{t}{\tau}\left(1 + \frac{\xi^2}{\pi^2} \right) \right), 
\label{eq:h0exp}
\eea
where $\xi := (\varepsilon_k - \mu)/(k_B T_\infty)$ and the mathematical form of $h(\varepsilon_k,t)$ is adopted from Ref.~\onlinecite{KaAl08}. Since it is an uneven function of $\xi$, the same deviation from equilibrium of both electrons and holes is assumed.  
As a possible generalization, different functions $h^{\pm}(\varepsilon_k,t) $ could be used for $\varepsilon_k > \mu$ positive and  $\varepsilon_k < \mu$ , i.e., for electrons and holes, as long as one makes sure that charge conservation is obeyed.  
The parameter $\tau$ is a material-specific quantity and typically lies in the range of a few picoseconds. 

For e-ph scattering we build on the assumption that it can be described by first-order perturbation theory using Fermi's Golden Rule, as has been previously employed in the literature. 
Both  for stimulated emission and for absorption of phonons by electrons, the corresponding rate will be proportional to $n_\omega(t)$, the population of the phonon state of frequency $\omega$.
This central quantity governing the rates will be a basic ingredient of the theory presented here. 
Moreover, this level of description allows us to make contact with the Migdal-Eliashberg function, a wide-spread concept to describe e-ph coupling, see e.g. Ref.~\onlinecite{Hofmann2009}. 
However, the presence of an excess contributions $h_k(t)$ in eq.~(\ref{eq:hk}) has the consequence that these extra carriers with energies only a few meV above or below the Fermi energy can contribute notably to the spontaneous emission of phonons. This is understandable because such electrons or holes can be long-lived (several picoseconds) with respect to e-e scattering, while their energy is still sufficiently high (comparable to the phonon energy scale of $\hbar \omega_D$, where $\omega_D$ is the Debye frequency) to allow for phonon emission. Hence, the role of the excess contribution for the overall energy transfer to the lattice needs to be considered. 

The structure of the paper is as follows: We  outline a kinetic theory based on rate equations that extend the established description of electronic relaxation to systems where deviations from a Fermi distribution function play a role. Subsequently, the rate constants are  specified based on microscopic descriptions of the scattering processes, in particular for screened Coulomb scattering in metals.  Generalization to semimetals \cite{Tomadin2013} or narrow-gap semiconductors seems to be possible using a different parameterization of the Coulomb scattering kernel. 
In Sect. III of the paper, results for a specific system, a film of few lead (Pb) atomic layers on a Si(111) substrate exposed to ultra-short laser pulses, are presented. A microscopic model for electron-phonon scattering, using an Eliashberg function from first-principles electronic structure calculations, is described. Solving the rate-equation theory for this specific case, we are in position to compare to experimental data from time-resolved photoemission spectroscopy. Finally, we conclude and discuss further applications of the theory presented.  

\section{Theory} 

\subsection{Rate-equation modelling}

As starting point, we sketch the overall structure of the theory we are aiming at. A detailed description how to calculate the rate constants will be given further down in the text. 
The model proposed here uses $T_e(t)$ and $n(t)$ as dynamic variables which stand for the electronic temperature and the occupation of phonon modes. 
The overall framework of our theory consists of a linear system of rate equations of the form
\be
 \frac{d}{dt}
\left(
\begin{array}{c}
T_{e}(t) \\
n(t) 
\end{array}
\right) = 
\left(
\begin{array}{ccccc}
0 & r_{01}  \\
r_{10}  & r_{11}  \\
\end{array}
\right)
\left(
\begin{array}{c}
T_{e}(t) \\
n(t) 
\end{array}
\right) + 
\left(
\begin{array}{c}
-\gamma+s_{e}(t) \\
s(t) 
\end{array}
\right) \, .
\label{eq:myReq}
\ee
The inhomogeneity on the very right of the differential eq.~(\ref{eq:myReq}) is the central aspect of this work and 
arises due to the presence of the excess distribution $h_k(t)$ of hot carriers. Indeed, 
we can consider the quantity $h_0 e^{-t/\tau}$ in eq.~(\ref{eq:h0exp}) as a 'small parameter'; in this sense, the theory presented here constitutes an extension of 'standard models' of electronic relaxation to the situation of a small additional deviation from quasi-equilibrium at temperature $T_e(t)$. 
When we put aside for the moment the inhomogeneities $s_e(t)$ and $s(t)$ in eq.~(\ref{eq:myReq}), the central rate constants defining the model are 
the matrix elements $r_{ij}$ and 
\be
\gamma = \frac{\pi \hbar g(E_F) k_B}{c_v} \lambda \langle \omega^2 \rangle \, .
\ee
With the usual expression for the specific heat of a free electron gas \cite{Ashcroft76}, $E_e =\frac{1}{2} c_v T_e^2$ and $c_v = g(E_F) k_B^2 \pi^2/3$ 
with the electronic density of states at the Fermi energy given by $g(E_F)$, one obtains 
\be
\gamma = \frac{3 \hbar}{\pi k_B} \lambda \langle \omega^2 \rangle 
\label{eq:gamma-conventional}
\ee 
This is a well-known expression \cite{Allan87} for the phenomenological electron-phonon coupling constant $\gamma_T = \gamma/T_e$. 
It describes the 'rate of cooling' (in Kelvin/picosecond) by the lattice which is observed following an excitation of the electronic system. In this way, it is possible experimentally \cite{BrKa90} to extract information about the material-specific parameter $\lambda \langle \omega^2 \rangle$ (and ultimately about $\lambda$, the dimensionless so-called mass enhancement factor in the theory of conventional superconductors) from time-resolved pump-probe experiments. Note that eq.~(\ref{eq:gamma-conventional}) doesn't require any further materials-specific information such as, e.g., the density of states of electrons or phonons. 
From a theoretical viewpoint, the kinetic coefficients are linked to moments of the Eliashberg function $\alpha^2 F(\omega)$ by 
\be
\lambda \langle \omega^j \rangle =  2 \int_{0}^{\omega_D} \! d\omega
\, \alpha^2 F(\omega) \omega^{j-1} \, ,
\label{eq:lambda_om_j} 
\ee
and we will use this relation in conjunction with first-principles
calculations of the Eliashberg function presented in Section
\ref{sec:microscopic_input}.

Another crucial parameter is the rate at which the energy deposited in the system arrives in the lattice degrees of freedom, described by the kinetic coefficient
\be
r_{10} = \frac{\pi g(E_F) k_B}{\bar \omega} \lambda \langle \omega^2 \rangle \, ,
\ee
where $\bar \omega$ is an average (as detailed below) vibrational frequency of the modes relevant in e-ph coupling.

The theory can be readily extended to cover coupling to multiple groups of phonons, labeled by an index $i = 1, \ldots N$. 
We use a vector notation for the occupation numbers $n_i(t)$ of these phonon modes, and for their couplings $r_{j0}$ to the electronic temperature,  
$$
\vec n = 
\left(
\begin{array}{c}
n_1(t) \\
\vdots \\
n_N(t) 
\end{array}
\right), \qquad 
\vec r = 
\left(
\begin{array}{c}
r_{10}(t) \\
\vdots \\
r_{N0}(t) 
\end{array}
\right)
$$
in conjunction with a matrix $\mathbf{R} = (r_{jk})$ of kinetic coefficients, $j,k= 1, \ldots N$. 
The diagonal elements  $r_{jj}, \,  j=1,N$  describe the absorption of a phonon that transfers its energy to the electronic system, while  
the off-diagonal elements could be used to describe direct coupling
among different phonons or groups of phonons, e.g. due to anharmonic
effects of the lattice. For the time scales of interest in the present
work, the latter couplings can be disregarded. 
The role of phonon-phonon coupling in general has been investigated
  theoretically in Ref.~\onlinecite{MaCa17}. For the specific material Pb we study
  later in this paper, the time scale for the mode conversion from
  perpendicular to parallel modes and vice versa  
has been determined to be $\sim 30$ps and $\sim 100$ps, respectively, by classical molecular dynamics
simulations~\cite{SaKr13}. 
Yet it should be noted that indirect energy  transfer between different phonon
modes via the electronic system is possible in the present theory.
In addition, the line vector $\rho = (r_{01}, \ldots r_{0N})$ contains the coupling constants of the electronic temperature to the occupation of phonon modes. 
This gives us the coupled multi-temperature rate equations 
\be
 \frac{d}{dt}
\left(
\begin{array}{c}
T_{e}(t) \\
\vec n(t) 
\end{array}
\right) = 
\left(
\begin{array}{ccccc}
0 & \rho   \\
\vec r  & \mathbf{R}  \\
\end{array}
\right)
\left(
\begin{array}{c}
T_{e}(t) \\
\vec n(t) 
\end{array}
\right) + 
\left(
\begin{array}{c}
-\gamma+s_{e}(t) \\
\vec s(t) 
\end{array}
\right) \, 
\label{eq:myREQ}
\ee
that are at the heart of our theory. 
For the new ingredients, the inhomogeneity $\vec s(t)$ with components $s_i(t)$, we will show in the appendix that it consists of a sum of terms linear and quadratic in the small parameter $h_0 e^{-t/\tau}$, 
\be
s_i(t) = \frac{\pi  g(E_F)}{\bar \omega_i} \sum_{m=1}^{2} \sigma_{i,m}(t) \bigl( h_0 e^{-t/\tau} \bigr)^m \, ,
\label{eq:s_i}
\ee
while the coefficients $\sigma_{i,m}(t) $ can again be determined from the suitably defined moments of the Eliashberg function.

It is instructive to see how to regain from this theory the  
standard formulations of a two-temperature or multi-temperature model. 
 For phonons that can be described by classical statistics, i.e. for sufficiently high occupation numbers, one may approximate $n_i = k_B T_i/(\hbar \bar \omega_i)$. 
 In this way, we introduce an effective  'lattice temperature' $T_i(t)$ associated with a certain group of phonons with characteristic frequency centered around $\bar \omega_i$. 
 
 For the two-temperature and multi-temperature models, it is the difference $T_e(t) -T_i(t)$ that acts as driving force toward equilibration. 
 In order to cast our mathematical formalism into this framework, one has to introduce a 'constant'  
 $\gamma/T_e$ 
(obviously depending on $T_e$, but for considering the approach to equilibrium, $T_e \approx T_{\infty}$, this dependence may be suppressed). 
If, in addition, the inhomogeneities $s_e(t) $ and $s_i(t)$ in eq.~(\ref{eq:myREQ}) are dropped, one obtains
\be
 \frac{d}{dt}
\left(
\begin{array}{c}
T_{e}(t) \\
T_i(t) 
\end{array}
\right) = 
\left(
\begin{array}{ccccc}
-\gamma/T_e & {\tilde r}_{0i}  \\ {\tilde r}_{i0}  & {\tilde r}_{ii}  \\
\end{array}
\right)
\left(
\begin{array}{c}
T_{e}(t) \\
T_i(t) 
\end{array}
\right) \, ,
\ee
which is an example of the familiar two-temperature model. 
In such a model, the largest eigenvalue determines the unique time scale for the relaxation towards equilibrium of both the electronic and lattice degrees of freedom.

As mentioned already in the Introduction, there are important differences between multi-temperature models and the theory presented here:
The additional term $s_e(T)$ on the right hand side of the equation (\ref{eq:myREQ}) allows for a time dependence different from a linear decay of the electronic energy, (or a quadratic decay of the temperature,) as typically proposed by standard multi-temperature models.
Even if the excess quantities $s_e(t)$ and $\vec s(t)$ are neglected and the classical limit is taken, we note that the electron-phonon coupling terms ${\tilde r}_{0i}$ involve higher moments of the Eliashberg function, proportional to $\lambda \langle \omega^3 \rangle$, rather than  $\lambda \langle \omega^2 \rangle$ as in standard models. As a consequence, the high-energy part of the Eliashberg function will gain higher weight in calculating electron-phonon coupling parameters. 

\subsection{Electron-electron scattering only}
  The total energy of the electronic system consists of two parts, a contribution from the quasi-thermal distribution and from the excess carriers, respectively, 
\bea
E_e &=& E_{\rm thermal} + E_{\rm excess} = \nonumber \\
     &=& \frac{1}{2} c_v T_e^2 + \int \! d\varepsilon \, \varepsilon \, g(\varepsilon) \, h(\varepsilon, t)  
     \label{eq:elecEne}
\eea
We assume that the non-equilibrium electron distribution can be cast into the form given by eq.~(\ref{eq:hk})
for all times. The effect of the density of states can be absorbed in the prefactor $h_0$ in eq.~(\ref{eq:h0exp}) 
which is determined by the initially deposited energy of the laser pulse. 
The parameter $\tau$ is a characteristic of the material. 
Since e-e scattering in the initial phase is relatively fast, we assume that it will instantaneously re-establish the distribution in eq.~(\ref{eq:hk}), even if other mechanisms, e.g. e-ph scattering, tend to induce deviations. 
Moreover, since e-e scattering is energy conserving, the total electronic energy (in the assumed absence of e-ph scattering) must be constant: 
 \bea
 \lefteqn{\frac{dE_e}{dt} = 0 =  \frac{d}{dt} \Bigl( \frac{1}{2} c_v T_e(t)^2 +}  \nonumber \\
 &  + & h_0 g(E_F) (k_B T_\infty)^2 \frac{\pi^{7/2}}{2} \left(\frac{\tau}{t} \right)^{3/2}  e^{-t/\tau} \Bigr)
 \eea
From this we derive 
 \be
s_e(T_e(t), t) = h_0 \frac{3 \pi^{3/2} T^2_{\infty} }{2 \tau T_e(t) }
\left( 1 + \frac{3 \tau}{2 t} \right) \left( \frac{\tau}{t}
\right)^{3/2} e^{-t/\tau} 
\label{eq:s_e}
\ee
 as the mathematical form of the inhomogeneity in the rate eq.~(\ref{eq:myREQ}). 
In physics terms, this expression describes the creation or annihilation of excess electron-hole pairs by fast Coulomb scattering during relaxation of the electronic system to the final equilibrium temperature.
 
 \subsection{Electron-phonon scattering}

The hot electronic system loses energy to the lattice by electron-phonon coupling that we choose to describe in quantum-mechanical first-order perturbation theory.  
Following the work of Allen \cite{Allan87}, the energy in the electronic systems decays as 
\begin{equation}
\frac{dE_e}{dt} = -\frac{2 \pi}{\hbar N_c} \sum_{k,k'} \hbar \omega_q |M^q_{k k'}|^2 S(k,k') \delta(\varepsilon_k - \varepsilon_{k'} + \hbar \omega_q)
\label{eq:GoldenRule}
\end{equation}
where the matrix element is normalized to the unit cell, and $N_c$ counts the unit cells in the sample. 
In our notation we have assumed a metal with a single band crossing the Fermi energy (containing both electrons and holes) and thus suppressed the band index. However, it is possible to extend the theory to a metal with multiple bands or even to a semimetal or small-gap semiconductor where electrons and holes reside in different bands.  
Both momentum and energy conservation must be satisfied in electron phonon scattering. Hence the crystal momentum of the phonon is $q = k - k'$. The energy of the phonon branch $\omega_q$ must match the energy difference between initial and final state of the electron, $\hbar \omega_q = \varepsilon_{k'} - \varepsilon_{k}$.  
The scattering kernel $S(k,k')$ is defined\cite{Allan87,LiZh08} as
$$
S(k,k') = \bigl( f_k(t) - f_{k'}(t) \bigr) n_{\omega_q}(t)  - \bigl(1-f_k(t) \bigr) f_{k'}(t)
$$
The first term describes both absorption of a phonon by an electron and induced emission of phonons, and hence is proportional to the number $n_{\omega_q}(t)$ of phonons present at any instant in time.
The second term describes spontaneous emission. This process requires electronic  occupation of the state with quantum number $k'$, and a 'hole' in state $k$, i.e., the final electronic state must be unoccupied. 

Rather than working with matrix elements and discrete transitions, it is more convenient to work with the Eliashberg function and continuous integration over energy variables for the metallic bands. This approximation is justified if all relevant scattering processes take place in a small energy interval around the Fermi energy, such that the Eliashberg function is a valid description in the whole energy range. 
Moreover, the electronic density of states is required to show little variation around $E_F$, else this variation can be accounted for by an additional factor multiplying $\alpha^2 F(\omega)$. (see e.g. Ref.~\onlinecite{Wang94,PeIn13,Waldecker16}). The increase in lattice energy $E_{\rm lat}$ 
is (apart from the sign) equal to the decrease in electronic energy given by eq.~(\ref{eq:GoldenRule}),  
\be
\frac{dE_{\rm lat}(t)}{dt} = - \frac{dE_e(t)}{dt}
\ee 
and hence can be expressed as  
\begin{eqnarray}
\frac{dE_{\rm lat}(t)}{dt} &=&  \inte_{0}^{\omega_D} \! \! d\omega \,\hbar \omega \, \alpha^2 F(\omega) \inte_{-\infty}^{\infty} \! \! d\varepsilon 
\inte_{-\infty}^{\infty} \! \! d\varepsilon' \,  
\delta(\varepsilon - \varepsilon' + \hbar \omega) \nonumber \\ 
& \times & 2 \pi N_c g(E_f)  \bigl[ (f(\varepsilon) - f(\varepsilon') ) n(\omega, T_i) - \nonumber \\
& - & f(\varepsilon') (1-f(\varepsilon))\bigr]  
\label{eq:energy-loss}
\end{eqnarray}
Here the Eliashberg function $\alpha^2 F(\omega)$ carries the information both about the size of the matrix element in eq.~(\ref{eq:GoldenRule}) as well as about the electronic density of states $g(\varepsilon)$.
We need to insert for the electronic distribution functions $f(\varepsilon,T_e(t))$ the sum of both the quasi-equilibrium term (at temperature $T_e$) and the excess  term $h(\varepsilon,t)$.
Electron-hole symmetry has been assumed for the excess term. 
Therefore, as long as only the contribution proportional to $f(\varepsilon) - f(\varepsilon')$ (first term in the square bracket of eq.~(\ref{eq:energy-loss}), describing phonon absorption and induced emission) is considered, $f$ can be replaced by $f^{(0)}$ without making an error; one can show that any terms being linear in $h(\varepsilon,t)$ 
do not contribute to the first term in the integral. 
One of the energy integrals in eq.~(\ref{eq:energy-loss}) can be carried out easily due to the energy conserving $\delta$-function. For the remaining energy integral over the terms containing $f^{(0)}$, one uses the well-known Sommerfeld expansion. The effective width of $-\partial f^{(0)}/\partial \varepsilon$ introduces an additional factor $\hbar \omega$. Finally the integration over $\omega$ needs to be carried out. 

Often it is convenient to consolidate the lattice vibrations into groups rather than following each $n_{\omega}(t)$ individually. In particular, this is appropriate if several phonons couple to the same thermal bath. In this case, one may switch to lattice temperatures $T_i(t)$ as new dynamic variables that are related to the populations via Bose-Einstein distribution functions $n_{\omega}(t) = n_B(\omega,T_i(t)) = [\exp(\hbar \omega/k_B T_i(t))-1]^{-1}$. Depending on modeling demands, the grouping may also follow other principles, e.g. transversal versus longitudinal polarization of the phonon. Here, however, we use the frequency as distinguishing feature, i.e. we group together all phonons in a frequency interval $[\Omega_{i-1}, \Omega_{i}], \; i=1, \ldots N$. We can define a weighted-average phonon frequency in each group by
\be
\bar \omega_i =  \frac{ \int_{\Omega_{i-1}}^{\Omega_i} \! d\omega \, \omega G(\omega) }{ \int_{\Omega_{i-1}}^{\Omega_i} \! d\omega \, G(\omega) } \,
\ee
where $G(\omega)$ is the density of states of the phonons. 
Similarly, the moments of the Eliashberg function are restricted to the relevant energy ranges,
\be
\lambda_{i}\langle \omega^j \rangle =  2 \int_{\Omega_{i-1}}^{\Omega_i} \! d\omega \, \alpha^2 F(\omega) \omega^{j-1} \, .
\label{eq:lambda_i}
\ee

Evaluating the integrals in eq.~(\ref{eq:energy-loss}) using the above expressions, 
the rate constants entering the matrix of the rate equation (\ref{eq:myREQ}) can be obtained. 
For relating the spontaneous emission of phonons to the electronic temperature $T_e$, it is required to employ the additional approximation (see Ref.~\onlinecite{MaCa17}, eq.~(A20))
\bea
 \bigl(1-f^{(0)}_k(t) \bigr) f^{(0)}_{k'}(t) &=& \bigl(f^{(0)}_k - f^{(0)}_{k'} \bigr) n_B(\omega_q, T_e(t)) \nonumber \\
 & \approx & -k_B T_e(t) \frac{\partial f^{(0)} }{\partial \varepsilon_k}
\label{eq:nB_Te}
\eea
Thus  one obtains 
\be
r_{i0} = \frac{ \pi g(E_F) k_B}{\bar \omega_i}  \lambda_i \langle \omega^2 \rangle,  
\mbox{ and } 
\gamma = \sum_{i=1}^N r_{i0} \, .
\label{eq:r_i0}
\ee
When obtaining the kinetic coefficients responsible for induced emission and absorption, we couple the electronic temperature $T_e(t)$ directly to the occupation numbers $n_i(t)$. 
This leaves us with the additional factor $\omega$ inside the integral of eq.~(\ref{eq:energy-loss}), and we thus obtain
\be
r_{0i} =  \frac{ \pi \hbar^2 g(E_F)}{c_v T_e}  \lambda_i \langle \omega^3 \rangle 
 =  \frac{ 3 \hbar^2}{\pi k_B^2 T_e}  \lambda_i \langle \omega^3 \rangle 
\label{eq:r_0i}
\ee
and 
\be
r_{ii} =  -\frac{ \pi \hbar g(E_F)}{\bar \omega_i}  \lambda_i \langle \omega^3 \rangle \, .
\label{eq:r_ii}
\ee

In addition to absorption and induced emission considered so far, including {\em spontaneous} emission of phonons by the hot carriers requires  
consideration of the non-equilibrium distributions $h_k(t)$ in the kernel $S(k,k')$, 
\begin{eqnarray}
S(k,k') &= \bigl( f^{(0)}_k(t) - f^{(0)}_{k'}(t) \bigr)  \bigl( n_{\omega_q}(t) - n_B(\omega_q, T_e(t)) \bigr) - \nonumber \\ 
&   -\bigl(1-f^{(0)}_k(t) \bigr) h_{k'}(t) + h_k(t) f^{(0)}_{k'}(t) + h_k(t) h_{k'}(t) \, . \nonumber
\end{eqnarray}
This leads to additional terms in the coupled system of rate equations that depend 
explicitly on time.
For the specific model of $h_k(t)$ provided in eq.~(\ref{eq:h0exp}) and for the grouping of the phonons according to their frequency ranges, the time-dependent inhomogeneities in the rate equation have been worked out in the appendix. 
The result has the form of eq.~(\ref{eq:s_i}) with the coefficients 
\bea
\sigma_{i,1}(t) &=&  \lambda_i \langle \omega^2 \rangle \frac{\pi^2 \tau k_B T_\infty}{4t}  K_0(t)- \nonumber \\
&-&  \lambda_i \langle \omega^4 \rangle \Bigl( \frac{\hbar^2 K_0(t)}{4 k_B T_\infty} - \frac{ \hbar^2 T_e(t)^2 }{2 k_B T_\infty^3}  K_2(t)  \Bigr) \,
\label{eq:sigma1} \\
\sigma_{i,2}(t) &=&  \lambda_i \langle \omega^2 \rangle k_B T_\infty \frac{\pi^{7/2}}{2^{5/2}} \left( \frac{\tau}{t} \right)^{3/2} - \nonumber \\
& - & \lambda_i \langle \omega^4 \rangle \frac{1}{k_B T_\infty} \frac{3\hbar^2 \pi^{3/2}}{2^{7/2}} \left( \frac{\tau}{t} \right)^{1/2}
\label{eq:sigma2}
\eea
where the remaining integrals to be calculated numerically are specified by
\be
K_j(t) = \int_{-\infty}^{\infty} \! d\eta \, \exp\left( -\frac{T_e(t)^2 t}{\pi^2 T_\infty^2 \tau} \eta^2 \right) \frac{\eta^j}{\cosh^2 \eta/2} \, .
\ee

Finally, we would like to point out the connection to multi-temperature models. 
If only the thermal contribution is retained, i.e., if one uses $f^{(0)}_k(T_e(t))$ instead of $f(\varepsilon,t)$ everywhere, one can make use of the identity (\ref{eq:nB_Te}) to simplify the scattering kernel to 
\begin{equation}
S^{(0)}(k,k') =  \bigl(f^{(0)}_k - f^{(0)}_{k'} \bigr)  \bigl( n_{\omega_q}(t) - n_B(\omega_q, T_e(t)) \bigr) \, .
\label{eq:clever}
\end{equation}
Thus $S^{(0)}$ reflects the balance between energy flows to and from the heat bath represented by the lattice. 
For small phonon energies $\hbar \omega_q$ and high temperatures, one can approximate the Bose-Einstein distribution function by its limit in classical statistics, and thus obtains
\begin{equation}
S^{(0)}(k,k') \approx  \bigl(f^{(0)}_k - f^{(0)}_{k'} \bigr) \frac{k_B(T_{\rm lat}(t) - T_e(t) )}{\hbar \omega_q}
\label{eq:classicalPhonons}
\end{equation}
This equation shows that the energy flow between the electron and phonon systems, at least for long times where both systems are in quasi-equilibrium at some temperature, 
is proportional to the temperature difference. This provides the microscopic foundation for the two-temperature, or likewise for multi-temperature models if the phonon system is divided into subsystems each with their own temperature. 
However, in our present theory we retain the $n_{\omega_q}(t)$ explicitly as dynamic variables. 
For the rightmost term in eq.~(\ref{eq:clever}), it is safe to use $n_B(\omega_q, T_e(t) ) \approx k_B T_e(t)/(\hbar \omega_q)$ since $T_e(t)$ is  usually much higher than the Debye temperature of the lattice.

\section{Application to time-resolved spectroscopy of P\lowercase{b} films}

Thin films of Pb on Si(111) can be grown at varying thickness that can be determined with monolayer (ML) precision. Both the electronic and vibrational structure of monolayer Pb films strongly depend on the film thickness. 
The band gap of Si(111) inhibits propagation of excitations from the Pb film into the substrate. 
Therefore the system Pb/Si(111) is an ideal probing ground to study microscopic theories of electron-phonon interactions and their implication for electronic relaxation experiments. In this work, we focus on a 5 ML Pb/Si(111) film.  

\begin{figure}
	\centering
	\includegraphics[width=0.75\linewidth]{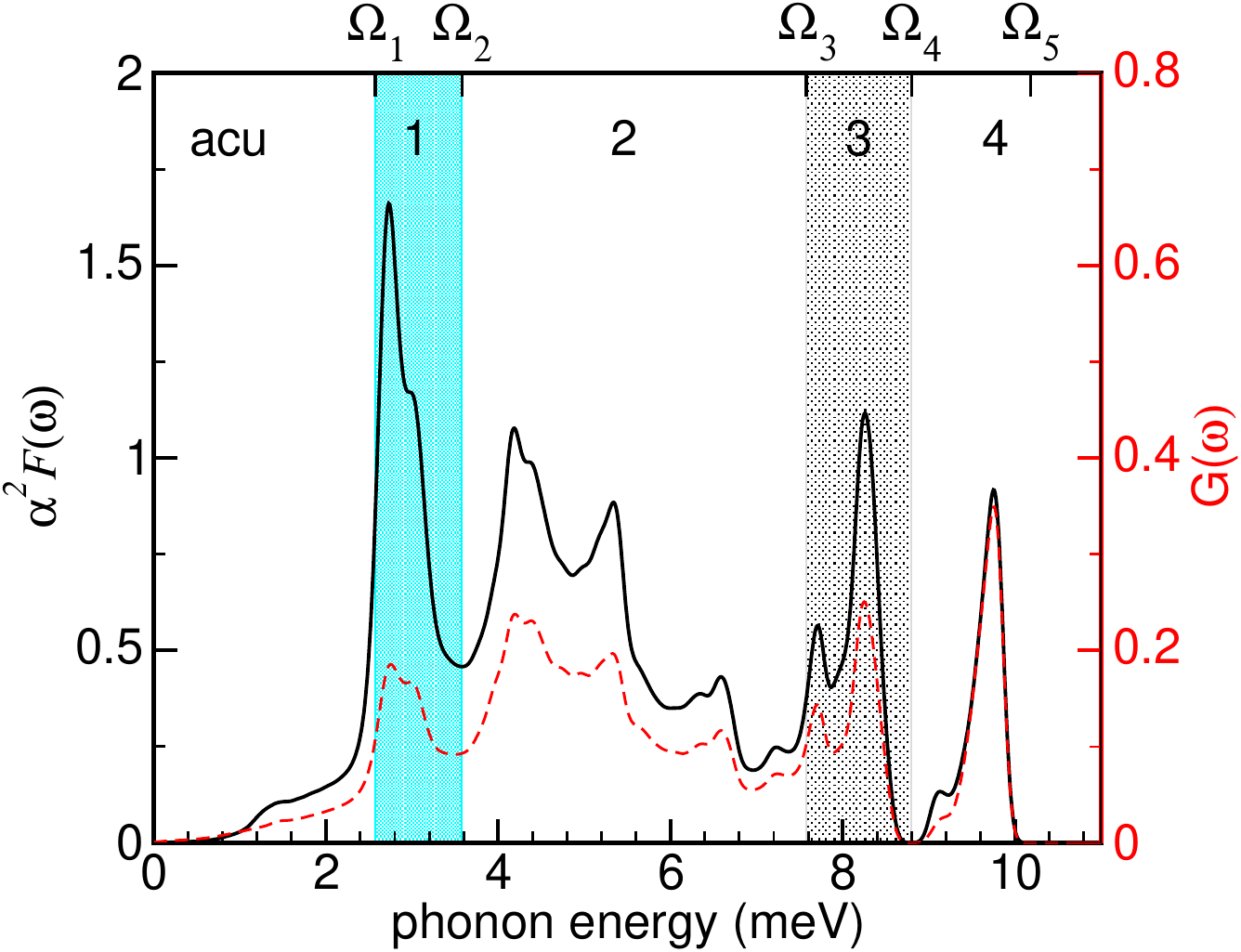} 
	\caption{Eliashberg function $\alpha^2 F(\omega)$ (full
          black line, left scale) and phonon density of states $G(\omega)$ (red
          dashed line, right scale) for a free-standing Pb film of
          thickness 5 monolayers from first-principles
            calculations similar to Ref.~\onlinecite{SkHe13}. 
          The frequency scale has been divided into the range of
          acoustic phonons plus four ranges describing vibrations of
          the Pb layers against each other, 
          marked by the labels $\Omega_i$. 
}
	\label{fig:Eliashberg}
\end{figure}

\subsection{Microscopic input} \label{sec:microscopic_input}

The Eliashberg function $\alpha^2F(\omega)$ and phonon density of states $G(\omega)$ of a free-standing film of Pb of 5 monolayer thickness have been calculated using density functional theory in the local density approximation \cite{HeLu71}. A mixed-basis pseudopotential approach \cite{Louie79} combining a plane-wave expansion with local functions is used to describe the valence electrons of Pb. The phonon properties and the electron-phonon interaction are obtained using a linear-response technique.\cite{HeBo99,Zein84,BaGi01} 
Spin-orbit coupling has been implemented as described in Ref.~\onlinecite{HeBo10}.
As was shown in Ref.~\onlinecite{SkHe13} the spin-orbit interaction not only has a strong effect on the phonon spectrum, but also leads to a strong enhancement of the electron-phonon coupling strength in thin Pb films.  
The calculated $\alpha^2F(\omega)$ and $G(\omega)$ are displayed in Fig.~\ref{fig:Eliashberg}.
The mass-enhancement parameter is obtained 
 as $\lambda = 2.04$ and $\lambda \langle \omega^2 \rangle = 46.2$~meV$^2/\hbar^2$ 
 by integrating the calculated Eliashberg function over the full frequency range.

Later we intend to couple phonons of similar character to a common thermal bath. 
Guided by the sharp maxima in the Eliashberg function in Fig.~\ref{fig:Eliashberg}, 
we therefore divide the energy range of phonons into five intervals: Below $\Omega_1$, there are only the acoustic phonon branches contributing \cite{BeSk18}, and these couple only weakly to the electrons. The higher-lying intervals involve motion of neighboring layers or neighboring Pb atoms relative to each other.  
There is strong coupling between the electrons (in particular those in the quantum well state near $E_F$) and phonon modes normal to the film, in which both the top and bottom layer of the film are moving in phase (frequency interval $[\Omega_1,\Omega_2]$) or out-of-phase 
 (frequency interval $[\Omega_3,\Omega_4]$), both shown as shaded areas in Fig.~\ref{fig:Eliashberg}.  
In addition, there is somewhat weaker coupling to phonons of mixed character in the intermediate frequency range $[\Omega_2,\Omega_3]$. 
The highest peak in the frequency range $[\Omega_4,\Omega_5]$ stems from surface phonons involving the in-plane motion of Pb atoms. 
This characterization of the phonons remains essentially unchanged
when considering more realistic atomic structures, e.g. a 5~ML Pb film
grown on Si(111) and DFT calculations of this system employing a
$(\sqrt{3} \times \sqrt{3})$ supercell~\cite{ZaKr17}. 

\subsection{Experimental results}

A Pb film of 5~ML thickness has been prepared on the Si(111) surface, as described previously \cite{KiBo07,KiBo08}. 
Time-resolved photoemission spectroscopy is performed on this sample using femtosecond laser pulses from a Ti:sapphire laser producing IR pulses at 1.5~eV photon energy. The temporal duration of the pump pulse was 50~fs. The overall absorbed fluence (Pb film and underlying Si substrate) has been determined to 1.1 mJ/cm$^2$. 

\begin{figure*}
	\centering
	\includegraphics[width=0.7\linewidth]{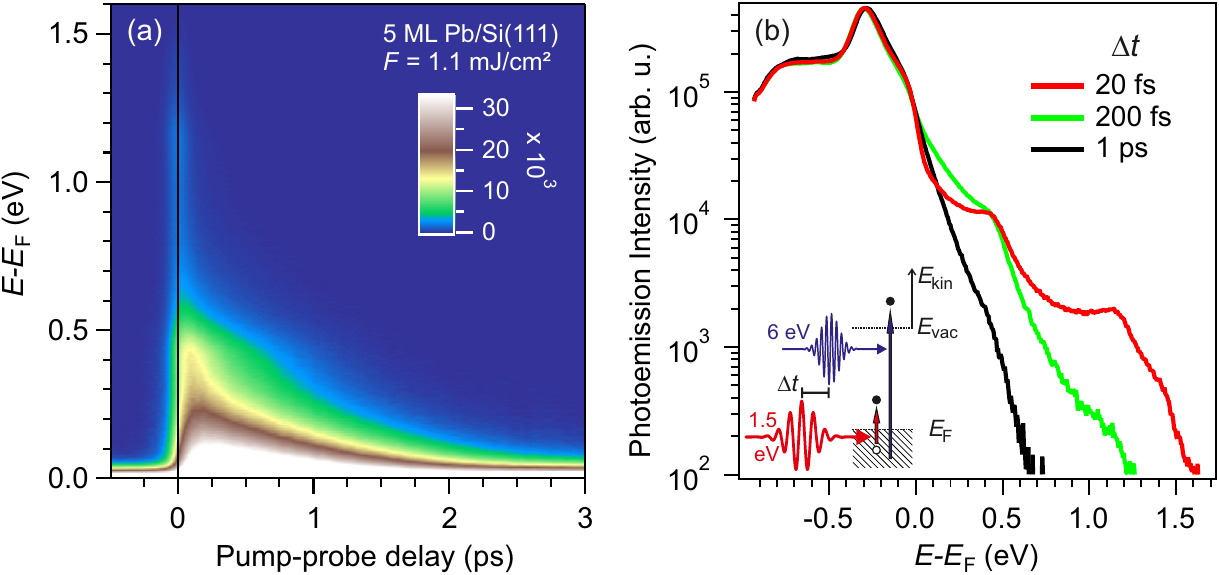} 
	\caption{(a) Time-dependent photoemission intensity at energies above the Fermi energy $E_\mathrm{F}$ in a false color representation obtained in normal emission geometry in a pump-probe experiment with pump and probe photon energies of 1.5 and 6.0~eV, respectively, on 5~ML Pb/Si(111). The absorbed pump fluence $F$ is indicated. Data were taken at a sample temperature of 80~K. (b) Selected photoemission spectra on a logarithmic intensity scale extracted from data in panel (a) at the indicated time delay $\Delta t$. The inset depicts a scheme of the pump-probe experiment.
}
	\label{fig:TRPE}
\end{figure*}

Photoemission spectra recorded with a frequency-quadrupled probe pulse of 6~eV photon energy and 100~fs duration at various pump-probe delays are shown in Fig.~\ref{fig:TRPE}. The experimental details are similar to those reported in Ref.~\onlinecite{KiRe10,ReKi12}. Fig.~\ref{fig:TRPE}(a) depicts the detected photoemission intensity in normal emission as a function of energy above $E_\mathrm{F}$ and as a function of time delay $\Delta t$. Shortly before and after $\Delta t = 0$ photoemission intensity is recorded almost up to the pump photon energy within a narrow time window of 100~fs which is determined by the pump pulse photon energy and the probe pulse duration. Relaxation dynamics appears at a first glance to occur at energies below 0.7~eV. Panel (b) shows the identical data as spectra on a logarithmic intensity scale at selected time delays. Here, excitations up to 1.5~eV are clearly identified and the relaxation proceeds with increasing $\Delta t$ over all energies. The three peaks in these spectra are caused by an enhanced electron density of states and originate from formation of quantum well states due to confinement of the $6 p_z$ wave function to the film \cite{KiBo08,KiRe10}. The peak at $E-E_\mathrm{F}=-0.3$~eV is the highest occupied quantum well state, the peak at $E-E_\mathrm{F}=0.4$~eV originates from a fractional 6~ML coverage within the illuminated sample surface \cite{KiRe10}, and the peak at $E-E_\mathrm{F}=1.1$~eV is the first unoccupied quantum well state at resonance with an optical excitation by pump photons with the occupied quantum well state. These peaks are populated according to the transient electron distribution and, at first glance, these spectra in Fig.~\ref{fig:TRPE} show a thermal distribution of electrons. Their temperature $T_e(t)$ decays on a time scale of approximately 1~ps. 

\begin{figure}
\begin{tabular}{lc}
	a) & 	\\ & \includegraphics[width=0.7\linewidth]{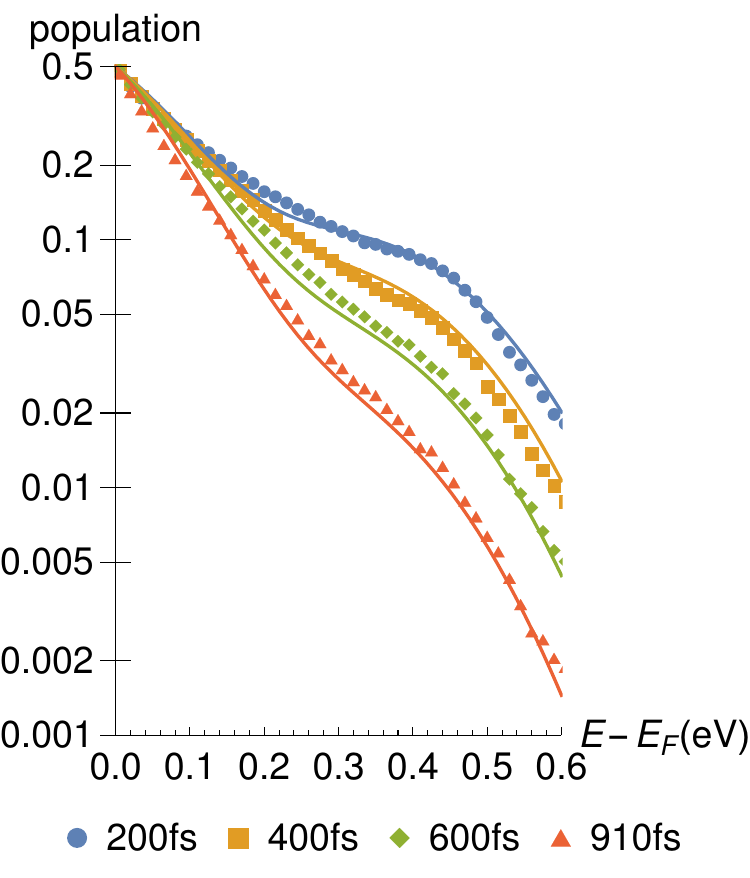} \\ 
	b) & \\ & \includegraphics[width=0.7\linewidth]{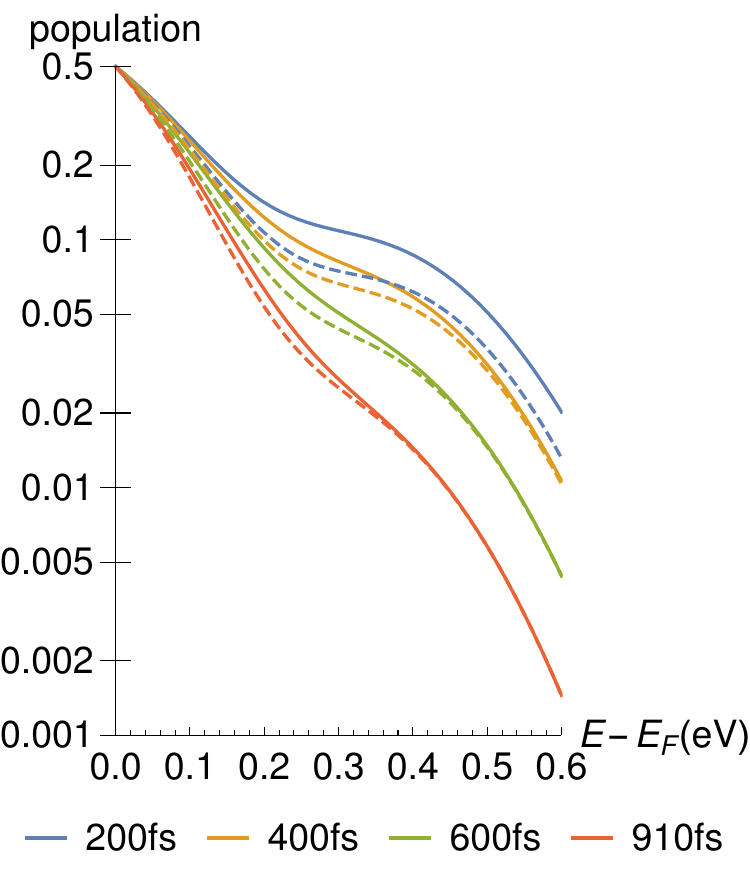}
\end{tabular}
	\caption{a) Logarithmic plot of the time-dependent population
          of excited states from time-resolved photoemission (symbols)
          and from a fit (lines) with eq.~(\ref{eq:fitting}) and
          (\ref{eq:h0exp}) using $h_0 = 0.005$ and $\tau = 6$ps.
	b) Comparison of the complete fit (full lines) with a fit by a
	thermal distribution $f^{(0)}(T_e(t))$ (dashed lines) in which
	the electronic temperature has been matched at the Fermi
	energy. In both cases, the peak in the electronic density of states at 0.55~eV has been considered.
}
	\label{fig:expSpectra}
\end{figure}

To investigate the electronic distribution near $E_F$ in further detail, we zoom into a small part of the data and normalize the signal at $E_F$ to 1/2, the value of the Fermi function at $E_F$. The symbols plotted in Fig.~\ref{fig:expSpectra}(a) show these experimental data. 
When fitting the spectra, we include the above-mentioned sample-specific density-of states effect by an asymmetric  Gaussian peak added to a smooth (for practical purposes constant) density of states $g(E_F)$. 
The fitting function is 
\be
g(E) \times \left( \frac{1}{\exp\bigl( \frac{E-E_F}{k_B T_e(t)} \bigr) +1} + h(E,t) \right)
\label{eq:fitting}
\ee
with $h(E,t)$ given by eq.~(\ref{eq:h0exp}) and  
\be
g(E) = g(E_F) \left( 1 + (E - E_F) \, g_0 \, e^{-[(E - E_{\rm peak})/w]^2} \right)
\ee
with $g_0 = 27$, $E_{\rm peak} = 0.55$eV, and $w=0.2$~eV.  
The main part of the fitted distribution comes from a thermal distribution with a temperature $T_e(t)$ fitted to the slope of the measured distribution functions for each time $t$. 
We note that a thermal distribution alone is not sufficient to reproduce the measured spectra.  
In particular at energies  0.1 -- 0.2 eV above $E_F$ it falls off too quickly to properly describe the data. 
The fit is much improved by using the distribution from eq.~(\ref{eq:h0exp}) taking into account the excess non-thermal electrons. The parameter $h_0$ is chosen as 0.005. 
This amounts to about 1\% of the electrons belonging to the non-thermal component. 

\subsection{Simulations}

Next we solve the system of rate equations (\ref{eq:myREQ}) numerically for the specific initial conditions at time $t_0=70$~fs. 
The lattice vibrations of the Pb film have been described by five different phonon groups, each with a separate time-dependent occupation $n_i(t)$. 
Both for $T_e(t_0)$ and for the $n_i(t_0)$ we assume thermal equilibrium at the sample temperature $T_0= 80$~K. 
All material parameters entering into the rate equation model have been obtained from first-principles calculations: 
The Eliashberg functions and its moments were taken from the DFT calculations described above. 
The density of states near the Fermi energy in bulk Pb was obtained from separate DFT calculations to be $2.7 \times 10^{22}$eV$^{-1}$cm$^{-3}$. For a $(1 \times 1)$ surface unit cell of a film with $N_c=5$ atoms, this yields $g(E_F) = 4.2 \times 10^{-3}$meV$^{-1}$ used for the simulations.  
The only parameter that could not be determined from first principles is the relaxation time $\tau$  due to e-e scattering, for which a value of $\tau =6$~ps was assumed. 
Since $\tau$ in our model describes the equilibration by electron-electron scattering, this value may appear rather high. 
However, since its purpose is the modeling of low-lying electronic excitations with an energy comparable to phonons (see the discussion in Ref.~\onlinecite{GrSp95}) and the electronic lifetime is known to diverge as $(E - E_F)^{-2}$,  a lifetime of several picoseconds seems reasonable.

\begin{figure}
	\centering
	\includegraphics[width=0.8\linewidth]{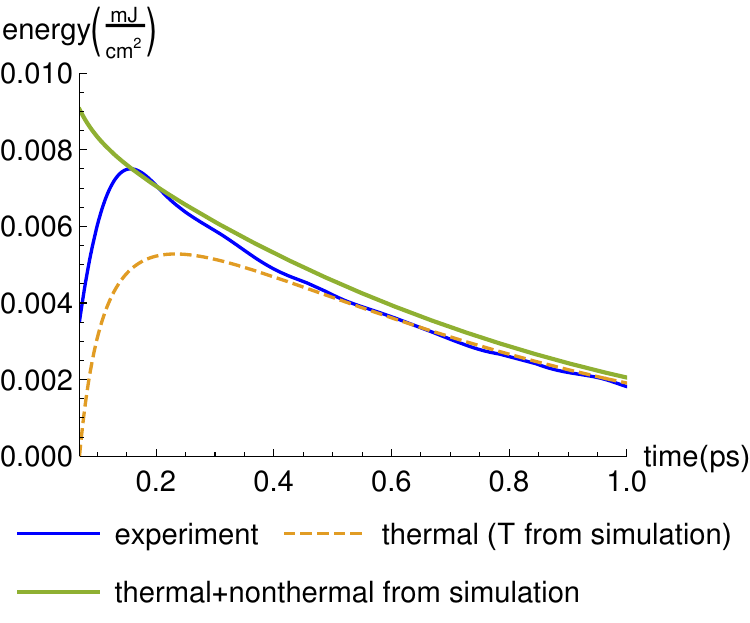} 
		\caption{Electronic energy obtained by integrating the experimental spectra 
		(blue line), thermal contribution 
                 using $T_e(t)$ from the simulation (dashed) and full simulation result including both thermal and nonthermal contributions (light green line).
	}
	\label{fig:energy}
\end{figure}

\begin{figure}
	\centering
	\includegraphics[width=\linewidth]{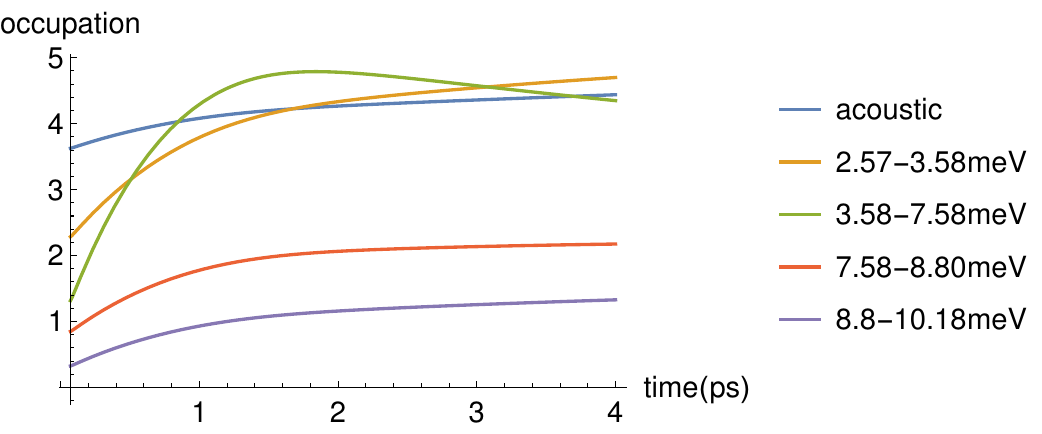}
	\caption{Average number of phonon quanta $n_i(t)$ in various ranges of the phonon spectrum, as obtained from the rate equations.
	}
	\label{fig:occu}
\end{figure}

\begin{figure}[th]
	\centering
	\includegraphics[width=\linewidth]{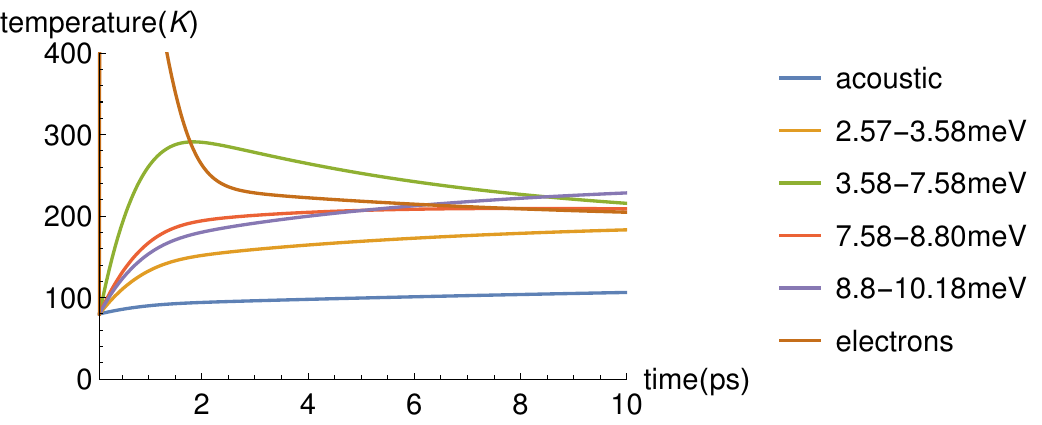}
	\caption{Effective temperatures $T_i(t)$ of various frequency ranges in the phonon spectrum and electronic temperature $T_e(t)$, as obtained from the rate equations.}
	\label{fig:phonontemp}
\end{figure}

The outcome of our simulations with the rate-equation model are presented in Fig.~\ref{fig:energy}, \ref{fig:occu} and \ref{fig:phonontemp}. 
We obtain the functions $T_e(t)$ and $n_i(t)$ from a numerical solution of the system of rate equations (\ref{eq:myREQ}) using  {\sc Mathematica12}. 
The microscopic description enters into (\ref{eq:myREQ}) via the
kinetic coefficients obtained from eq.s (\ref{eq:r_i0}), (\ref{eq:r_0i}),
(\ref{eq:r_ii}), (\ref{eq:s_i})  and (\ref{eq:s_e}).  
In these coefficients, the Eliashberg function enters via its moments
$\lambda \langle \omega^j \rangle$, see eq.~(\ref{eq:lambda_om_j}). 
In Fig.~\ref{fig:energy} we show the total electronic energy, eq.~(\ref{eq:elecEne}) 
with $T_e(t)$ as obtained from the simulations inserted.  
The first term in eq.~(\ref{eq:elecEne}), that is the thermal contribution determined by the temperature $T_e(t)$ alone, is also shown. 
As expected from our inclusion of the non-thermal distribution $h$  
we find that there is excess energy in the electronic system for 
several hundred femtoseconds. 
Up to this time, the e-e scattering processes have not yet managed to establish an equilibrium temperature in the electronic system. 
Subsequently, an exponential decay of the energy is observed with a time constant of 0.59~ps. 
 
Moreover, we compare the simulation results to the temporal evolution of the electronic energy as observed in experiment. 
This curve has been obtained by numerical integration of the measured photoemission spectra, weighted by a factor $E - E_F$, for all energies above $E_F$. Normalization by a constant factor has been used in order to 
show both the simulated and the measured energy curves in the same plot. Since the photoemission experiment was not calibrated, the  absolute number for the electronic energy could not be obtained from the measurements. 
The experimentally determined electronic energy shows an exponential decay with a time constant of 0.56~ps, i.e., the simulated and the measured decay are in good agreement. However, the temporal behavior near the experimentally observed maximum is reproduced in the simulations only if the excess contribution $h_k(t)$ is taken into account. 

It is noteworthy that reproducing the experimental results simply by a multi-temperature model, without any excess contribution, would require an initial electronic temperature $T_e(t_0) \sim 1200$~K. This is significantly higher than the maximum $T_e$ of 1000~K found in our analysis of the experimental data. Thus, the multi-temperature model tends to overestimate the initial electronic temperature. 

If the {\em ab initio}-determined
Eliashberg function is used to calculate the cooling rate directly via eq.~(\ref{eq:gamma-conventional}), 
780~K/ps is obtained for the present system. Thus, according to the
multi-temperature model, the electronic system would reach the final temperature $T_\infty$ 
after about $\tau_{\rm cool} = 1.5$~ps. This is in contrast to our simulation, where a
nearly exponential decay and a long-time tail of $T_e(t)$ are obtained.
If one attempted to extract the
parameter $\gamma_T$ from a direct (two-temperature model) fit to the
experimental data, the slope of 780~K/ps would be obtained at 460~K,
i.e., at about one half of the electronic temperature maximum.  We
provide this information because if one attempts to match experimental
data with the (unphysically linear) temperature decay taken from a
two-temperature model, a considerable uncertainty of the fit, and
hence an incorrect determination of the electron-phonon coupling
constant, may result, depending on the time at which the measured data
and the model are brought to match.

Fig.~\ref{fig:occu} shows the phonon population $n_i(t)$ 
in the different groups of phonons 
that is obtained by integrating the differential equations (\ref{eq:myREQ}) 
It is seen that the low-lying acoustic modes have a high population already in the initial stage, since they are easily excited thermally due to their low energy. However, the population of these modes increases only slightly because their coupling to the electronic system is weak.
The phonon modes between 2.57 and 3.58~meV, however, couple strongly to the electronic system, and we observe a steep rise of their population. 
Both findings can be understood already from Fig.~\ref{fig:Eliashberg} by comparing the Eliashberg function to the phonon density of states. The strong enhancement of the former in the range of 2.57 and 3.58~meV indicates a strong coupling of the electronic states to 'breathing modes' of the Pb film. 
A similar observation, selective coupling to high-lying phonons, has been observed if unoccupied quantum well states are initially populated by the laser pulse. Both the experiment \cite{ReKi12} and simulations~\cite{KrZa19} observe the population of a 2~THz (= 8.3meV) phonon in the Pb film subsequent to laser excitation of the electronic system.

Note that the acoustic modes still change their population even at times of 4~ps and beyond; they have not reached thermal equilibrium. 
From the electronic temperature $T_e(t)$ obtained by integrating the differential equations (\ref{eq:myREQ}) and  
plotted in Fig.~\ref{fig:phonontemp}, one can see a small, but notable amount of excess energy in the electronic system even at these long times. This excess energy of the electronic system is hardly visible in electronic spectra (cf. the small differences near $E_F$ in Fig.~\ref{fig:expSpectra}b). However, even electrons just a few meV above $E_F$ have sufficient energy to excite phonons. We also note that such a long-lived electronic excitation can be sustained since the phonons with energies between 2.57 and 3.58~meV that overshoot their equilibrium occupation already after 0.5~ps are able to feed energy back into the electronic system.

In summary, this leads us to the conclusion that strongly different time scales for the population of different phonon modes appears to be a rather general phenomenon that has been observed in a number of systems, including thin metallic films, e.g. Au~(Ref.~\onlinecite{ChTr16}) and Ni (Ref.~\onlinecite{MaCh19}) and two-dimensional materials, e.g. graphene~\cite{CaNo20} and phosphorene \cite{Seiler2021}. 
Thus, incomplete equilibration between multiple subsystems should routinely be taken into account in future modeling. 

\section{Conclusion}

In this paper we studied,  both theoretically and experimentally, the relaxation of photo-excited electrons in metals under the combined influence of electron-electron and electron-phonon scattering. Femtosecond time-resolved photoemission data for thin films of Pb are used to verify the theoretical modeling results.   
The lifetime of hot electrons w.r.t electron-electron scattering is strongly energy-dependent, with much enhanced lifetime of electrons close to the Fermi energy. For this reason, one {\em cannot} say that the time scales for relaxation due to e-e and e-ph scattering are strictly separated.  
While e-e scattering is dominant at short times $< 0.3$ps, some excess distribution of hot electrons that cannot be captured by a quasi-thermal Fermi-Dirac distribution survives even up to several ps. These hot carriers, albeit their energy being in the meV range, are still able to excite phonons, in particular in heavy metals such as Pb. 
Moreover, considering an electronic contribution including non-thermal carriers allows for an improved description of the electronic energy and electronic temperature at short times $< 0.3$ps.  
In the lattice degrees of freedom, we observed that the coupling to  
medium-range phonons is more efficient as compared to the coupling of low-lying acoustic phonons. 
Therefore the effective lattice temperature of these phonon modes rises quickly, while the acoustic modes lag behind and are not fully equilibrated even after several ps. The interplay between the non-thermal excess electrons and the phonon system can lead to an effective transfer of energy from  the quickly excited high-lying to the slow acoustic modes. 

\section{Acknowledgments}

Financial support  by Deutsche Forschungsgemeinschaft (DFG) within SFB 1242 "Non-equilibrium dynamics of condensed matter in the time domain", project number 278162697, and within the Emmy-Noether program (grant no. RE3977/1) is gratefully acknowledged. 
We acknowledge support by the Open Access Publication Fund of the University of Duisburg-Essen. 

\appendix
\section{Derivation of hot-electron contribution}
As a general starting point, the energy change of the lattice vibrations in the $i$th interval $[\Omega_{i-1}, \Omega_i]$ is given by
\begin{eqnarray}
\frac{dE_i(t)}{dt} &=& \inte_{\Omega_{i-1}}^{\Omega_i} \! \! \! d\omega \,\hbar \omega \, \alpha^2 F(\omega) \! \inte_{-\infty}^{\infty} \! \! d\varepsilon 
\inte_{-\infty}^{\infty} \! \! d\varepsilon' 
\delta(\varepsilon - \varepsilon' + \hbar \omega) \times \nonumber \\ 
& & 2 \pi N_c g(E_F)  \bigl[ (f(\varepsilon) - f(\varepsilon') ) n_\omega(t) - \nonumber \\ 
&-& f(\varepsilon') (1-f(\varepsilon))\bigr]  
\label{eq:start}
\end{eqnarray}
The electronic distribution function $f(\varepsilon)$ is the sum of the quasi-equilibrium term (at temperature $T_e(t)$) and the excess term $h(\varepsilon,t)$.
This gives rise to three sources of energy change in eq.~(\ref{eq:start}):
\bea
\frac{dE_i(t)}{dt} &=& 2 \pi N_c g(E_F) \int_{\Omega_{i-1}}^{\Omega_i} \! d\omega \,\hbar \omega  \, \alpha^2 F(\omega) \, \bigl(J_0(\omega,t) - \nonumber \\ 
&  & - J_1(\omega,t) + J_2(\omega,t) \bigr)
\label{eq:decomposition}
\eea
In the second term in the square bracket in \ref{eq:start}, describing spontaneous emission of phonons by the hot carriers, we need to split $f$ into $f^{(0)}$ and $h$ to proceed.  
The first term, however, representing phonon absorption and induced emission, can be brought to a simple form,  
\bea
J_0(\omega,t) &=& \inte_{-\infty}^{\infty} \! \! d\varepsilon \inte_{-\infty}^{\infty} \! \! d\varepsilon' \, \delta(\varepsilon - \varepsilon' + \hbar \omega) (f^{(0)}(\varepsilon) - f^{(0)}(\varepsilon')) \nonumber \\ 
& \times & \bigl( n_i(t)  - n_B(\omega, T_e(t)) \bigr) 
\eea
since the value of the integral remains unchanged when replacing  $f$ by $f^{(0)}$, as explained in the main text above. 
Note that the thermal distributions $f^{(0)}$ carry an implicit time dependence via their dependence on $T_e(t)$ which is not spelled out but taken into account in the calculations. 
Further simplification is achieved by using eq.~(\ref{eq:nB_Te}). 
The resulting terms, proportional to $n_i$ and $T_e$, respectively, are taken into account via the matrix elements $r_{jj}$ and $r_{0j}$ in the rate eq.~(\ref{eq:myREQ}).

The non-equilibrium terms $J_1(\omega,t)$ and $ J_2(\omega,t)$ in eq.~(\ref{eq:decomposition}) need special attention. 
We obtain a mixed term, linear in both $f^{(0)}$ and $h$, 
\bea
J_1(\omega,t) &=& \int_{-\infty}^{\infty} \! d\varepsilon \int_{-\infty}^{\infty} \! d\varepsilon' \, \delta(\varepsilon - \varepsilon' + \hbar \omega) \bigl[ h(\varepsilon',t) \nonumber \\
& \times &   (1-f^{(0)}(\varepsilon)) - h(\varepsilon,t) f^{(0)}(\varepsilon') \bigr]
\eea
and a term quadratic in $h$, 
$$
J_2(\omega,t) = \int_{-\infty}^{\infty} \! d\varepsilon \int_{-\infty}^{\infty} \! d\varepsilon' \, \delta(\varepsilon - \varepsilon' + \hbar \omega) h(\varepsilon',t) h(\varepsilon,t) \, .
$$
To carry out the double integral, a change of variables,
$$
\int_{-\infty}^{\infty} \! d\varepsilon \int_{-\infty}^{\infty} \! d\varepsilon' \mapsto \int_{-\infty}^{\infty} \! d\left({\varepsilon + \varepsilon' \over 2}\right) \int_{-\infty}^{\infty} \! d(\varepsilon - \varepsilon') 
$$
is useful. 
The second integration over the energy difference can be carried out exploiting the property of the $\delta$-function, and one obtains 
\bea
\lefteqn{\frac{dE_i(t)}{dt} = 2 \pi N_c g(E_F) \times} \nonumber  \\
& & \int_{\Omega_{i-1}}^{\Omega_i} \! d\omega \, \alpha^2 F(\omega) \, \hbar \omega \int_{-\infty}^{\infty} \! d\bar\varepsilon \, 
\sum_{m=0}^2 I_m(\bar\varepsilon,\omega,t)  \bigl( h_0 e^{-t/\tau}   \bigr)^m \nonumber
\eea
with $\bar \varepsilon = (\varepsilon + \varepsilon')/2$. 
For the remaining integral over $\bar \varepsilon$, 
various strategies are used: 
For the thermal part $I_0$, one can make use of the observation that the two energies 
$\varepsilon$ and $\varepsilon'$ must both fall into a narrow interval of width $k_B T_e$ to get any significant contribution. Hence one can use a Taylor expansion of the occupation number difference
$$
I_0(\bar\varepsilon,\omega,t) \approx  -k_B T_e(t) \left. \frac{\partial f^{(0)}(\bar\varepsilon, T_e(t) )}{\partial \bar \varepsilon}\right|_{\varepsilon=\bar\varepsilon}\,  
$$
which is, to leading order, independent of $\omega$, see also eq.~(\ref{eq:nB_Te}). 
For the mixed term 
we employ integration by parts. To this end, we define 
$$
H(\xi,t) = h_0 \frac{\pi^2 \tau}{2 t} k_B T_\infty  \exp\left( - \frac{t}{\tau}\left(1 + \frac{\xi^2}{\pi^2} \right) \right)
$$
such that $H'(\xi,t) = -h(\xi, t)$, the prime denoting differentiation with respect to energy. 
Using integration by parts, 
exploiting the fact that $h$ vanishes for $\xi \to \pm \infty$, $I_1$ can be written as
\bea
I_1(\bar\varepsilon,\omega,t) &=&  H(\bar \varepsilon - \hbar \omega/2,t) \left. \frac{ \partial f^{(0)}(\epsilon + \hbar \omega/2)}{\partial \epsilon}\right|_{\bar \varepsilon} - \nonumber  \\
&- & H(\bar \varepsilon + \hbar \omega/2,t) \left. \frac{ \partial f^{(0)}(\epsilon - \hbar \omega/2)}{\partial \epsilon}\right|_{\bar \varepsilon} \nonumber
\eea
Since the $\bar \varepsilon$-integration extends over the whole real axis, we have the freedom to introduce integration variables shifted differently by $\pm \hbar \omega/2$ for both terms, allowing us to split off a common prefactor, 
\bea
\lefteqn{I_1(\bar \varepsilon,\omega,t) =  \frac{\pi^2 \tau}{2 t} k_B T_\infty \exp\left( -\frac{\xi^2 t}{\pi^2 \tau} \right) \times } \nonumber \\ 
& & \left( \frac{ \partial f^{(0)}(\bar \epsilon + \hbar \omega)}{\partial \epsilon} - \frac{\partial f^{(0)}(\bar \epsilon - \hbar \omega)}{\partial \epsilon} \right)
\label{eq:I1}
\eea
Next, we use a Taylor expansion in the parameter $\beta = \hbar \omega/k_B T_\infty$,
$$
f'(\bar \varepsilon \pm \hbar \omega) = f'(\bar \varepsilon) \pm \hbar \omega f"(\bar \varepsilon) + {1 \over 2} (\hbar \omega)^2 f'''(\bar \varepsilon) + \ldots 
$$
The prime denotes derivation of the Fermi function with respect to energy. The term linear in $\hbar \omega$ multiplying the odd function $f"(\bar \varepsilon)$ drops out after being integrated over $\bar \varepsilon$. 
Thus, the integral $J_1(\omega,t)$ entering in eq.~(\ref{eq:decomposition}) has a leading term which does not depend on $\omega$, followed by a quadratic term in the expansion parameter $\beta$, and hence in $\omega$, 
$$
J_1(\omega,t) = \left[ - \frac{\pi^2 \tau}{4 t}  J_1^{(0)}(t) + \beta^2 J_1^{(2)}(t) + {\cal O}(\beta^4) \right] h_0 e^{-t/\tau}
$$
with
\begin{eqnarray}
J_1^{(0)}(t) &=& k_B T_\infty \int_{-\infty}^{\infty} \! d\eta \, \exp\left( -\frac{T_e(t)^2 t}{\pi^2 T^2_\infty \tau} \eta^2 \right) \frac{1}{\cosh^2 \eta/2} \nonumber \\ 
J_1^{(2)}(t) &=&  {1 \over 4} J_1^{(0)}(t) -  \frac{t}{2 \pi^2 \tau} \frac{T_e(t)^2}{T^2_\infty} \times \nonumber \\ 
& &\int_{-\infty}^{\infty} \! d\eta \, \exp\left( -\frac{T_e(t)^2 t}{\pi^2 T^2_\infty \tau} \eta^2 \right) \frac{\eta^2}{\cosh^2 \eta/2} \, . \nonumber
\end{eqnarray}
In the last integral, again integration by parts has been used.
\begin{widetext}
Finally we describe the evaluation of the 
term $J_2(\omega,t)$  quadratic in $h$. It has  has the explicit form
$$
h(k',t) h(k,t) = h_0^2 \, \exp\left( -\left(2 + \frac{(\hbar \omega)^2}{2 \pi^2 (k_B T_\infty)^2 } \right) \frac{t}{\tau}  \right) \frac{(\bar \varepsilon - E_F)^2 - (\hbar \omega/2)^2 }{(k_B T_\infty)^2}\exp\left( -\frac{2 t (\bar\varepsilon - E_F)^2}{\pi^2\tau (k_B T_\infty)^2} \right) 
$$
The Gaussian integral over the average energy $\bar \varepsilon$ can be carried out analytically, and one obtains 
$$
J_2(\omega,t) = h_0^2 e^{-2t/\tau} \frac{\pi^{7/2}}{2^{5/2}} \left( \frac{\tau}{t} \right)^{3/2} \left(1 - \frac{4 t}{\pi^2 \tau} \left( \frac{\hbar \omega}{2 k_B T_\infty} \right)^2 \right) \exp\left( -\frac{(\hbar \omega)^2 t}{2 \pi^2\tau (k_B T_\infty)^2} \right) \, .
$$
To be compatible with the approximation used for $J_1$, we finally expand the exponential in powers of $\beta = \hbar \omega/(k_B T_\infty)$ and obtain
$$
J_2(\omega,t) \approx  h_0^2 e^{-2t/\tau}  \frac{\pi^{7/2}}{2^{5/2}} \left( \frac{\tau}{t} \right)^{3/2} \left[ 
1 - \frac{6 t}{\pi^2 \tau} \left( \frac{\hbar \omega}{2 k_B T_\infty} \right)^2 \right] + {\cal O}( \beta^4 ) \, .
$$
\end{widetext}  
This allows us to evaluate all contributions via the moments of the Eliashberg function defined in eq.~(\ref{eq:lambda_i}). 
Eventually, the integration over $\omega$, involving the Eliashberg function, is carried out for all three contributions $J_m(\omega,t), \quad m=0,1,2$. 
The contributions from $J_1$ and $J_2$ show up in the coefficients $\sigma_{i,1}(t)$ and $\sigma_{i,2}(t)$ in eq.~(\ref{eq:sigma1}) and (\ref{eq:sigma2}), respectively.


\begin{thebibliography}{10}
	
	\bibitem{FaSt92}
	W.~S. Fann, R. Storz, H.~W.~K. Tom, and J. Bokor, Electron thermalization in
	gold, Phys. Rev. B {\bf 46},  13592  (1992).
	
	\bibitem{GrSp95}
	R.~H.~M. Groeneveld, R. Sprik, and A. Lagendijk, Femtosecond spectroscopy of
	electron-electron and electron-phonon energy relaxation in {Ag} and {Au},
	Phys. Rev. B {\bf 51},  11433   (1995).
	
	\bibitem{AnKa74}
	S.~I. Anisimov, B.~L. Kapeliovich, and T.~L. Perel'man, Electron emission from
	metal surfaces exposed to ultrashort laser pulses, Sov. Phys. JETP {\bf 39},
	375   (1974).
	
	\bibitem{Waldecker16}
	L. Waldecker, R. Bertoni, R. Ernstorfer, and J. Vorberger, Electron-Phonon
	Coupling and Energy Flow in a Simple Metal beyond the Two-Temperature
	Approximation, Phys. Rev. X {\bf 6},  021003  (2016).
	
	\bibitem{MaCa17}
	P. Maldonado, K. Carva, M. Flammer, and P.~M. Oppeneer, Theory of
	out-of-equilibrium ultrafast relaxation dynamics in metals, Phys. Rev. B {\bf
		96},  174439  (2017).
	
	\bibitem{LiLo04}
	M. Lisowski, P.~A. Loukakos, U. Bovensiepen, J. St{\"a}hler, G. Gahl, and M.
	Wolf, Ultra-fast dynamics of electron thermalization, cooling and transport
	effects in {Ru(001)}, Appl. Phys. A {\bf 78},  165   (2004).
	
	\bibitem{ObDe20}
	M. Obergfell and J. Demsar, Tracking the Time Evolution of the Electron
	Distribution Function in Copper by Femtosecond Broadband Optical
	Spectroscoscopy, Phys. Rev. Lett. {\bf 124},  037401  (2020).
	
	\bibitem{ReKa02}
	B. Rethfeld, A. Kaiser, M. Vicanek, and G. Simon, Ultrafast dynamics of
	nonequilibrium electrons in metals under femtosecond laser irradiation, Phys.
	Rev. B {\bf 65},  214303  (2002).
	
	\bibitem{MuRe14}
	B.~Y. Mueller and B. Rethfeld, Nonequilibrium electron--phonon coupling after
	ultrashort laser excitation of gold, Appl. Surf. Sci. {\bf 302},  24
	(2014).
	
	\bibitem{RaFr16}
	J.~D. Rameau, S. Freutel, A.~F. Kemper, M.~A. Sentef, J.~K. Freericks, I.
	Avigo, M. Ligges, L. Rettig, Y. Yoshida, H. Eisaki, J. Schneeloch, R.~D.
	Zhong, Z.~J. Xu, G.~D. Gu, P.~D. Johnson, and U. Bovensiepen, Energy
	dissipation from a correlated system driven out of equilibrium, Nat. Commun.
	{\bf 7},  13761  (2016).
	
	\bibitem{KemperFreericks2018}
	A.~F. Kemper, O. Abdurazakov, and J.~K. Freericks, General Principles for the
	Nonequilibrium Relaxation of Populations in Quantum Materials, Phys. Rev. X
	{\bf 8},  041009  (2018).
	
	\bibitem{CaNo20}
	F. Caruso, D. Novko, and C. Draxl, Photoemission signatures of nonequilibrium
	carrier dynamics from first principles, Phys. Rev. B {\bf 101},  035128
	(2020).
	
	\bibitem{CaNo22}
	F. Caruso and D. Novko, Ultrafast dynamics of electrons and phonons: from the
	two-temperature model to the time-dependent {Boltzmann} equation, Advances in
	Physics: X {\bf 7},  2095925  (2022).
	
	\bibitem{BaKa14}
	V.~V. Baranov and V.~V. Kabanov, Theory of electronic relaxation in a metal
	excited by an ultrashort optical pump, Phys. Rev. B {\bf 89},  125102
	(2014).
	
	\bibitem{KrZa19}
	P. Kratzer and M. Zahedifar, Relaxation of electrons in quantum-confined states
	in {Pb/Si}(111) thin films from master equation with first-principles derived
	rates, New J. Phys. {\bf 21},  123023  (2019).
	
	\bibitem{Tomadin2013}
	A. Tomadin, D. Brida, G. Cerullo, A.~C. Ferrari, and M. Polini, Nonequilibrium
	dynamics of photoexcited electrons in graphene: Collinear scattering, {Auger}
	processes, and the impact of screening, Phys. Rev. B {\bf 88},  035430
	(2013).
	
	\bibitem{KaAl08}
	V.~V. Kabanov and A.~S. Alexandrov, Electron relaxation in metals: Theory and
	exact analytical solution, Phys. Rev. B {\bf 78},  174514  (2008).
	
	\bibitem{Hofmann2009}
	P. Hofmann, I.~Y. Sklyadneva, E.~D.~L. Rienks, and E.~V. Chulkov,
	Electron-phonon coupling at surfaces and interfaces, New J. Phys. {\bf 11},
	125005  (2009).
	
	\bibitem{Ashcroft76}
	N.~W. Ashcroft and N.~D. Mermin, {\em Solid State Physics} (Saunders College,
	Philadelphia, 1976).
	
	\bibitem{Allan87}
	P.~B. Allen, Theory of Thermal Relaxation of Electrons in Metals, Phys. Rev.
	Lett. {\bf 59},  1460  (1987).
	
	\bibitem{BrKa90}
	S.~D. Brorson, A. Kazeroonian, J.~S. Moodera, D.~W. Face, T.~K. Cheng, E.~P.
	Ippen, M.~S. Dresselhaus, and G. Dresselhaus, Femtosecond Room-Temperature
	Measurement of the Electron-Phonon Coupling Constant $\lambda$ in Metallic
	Superconductors, Phys. Rev. Lett. {\bf 64},  2172  (1990).
	
	\bibitem{SaKr13}
	S. Sakong, P. Kratzer, S. Wall, A. Kalus, and M. {Horn-von Hoegen}, Mode
	conversion and long-lived vibrational modes in lead monolayer on silicon(111)
	after fs-laser excitation: a molecular dynamics simulation, Phys. Rev. B {\bf
		88},  115419  (2013).
	
	\bibitem{LiZh08}
	Z. Lin, L.~V. Zhigilei, and V. Celli, Electron-Phonon Coupling and Electropn
	Heat Capacity of Metals under Conditions of Strong Electron-Phonon
	Nonequilibrium, Phys. Rev. B {\bf 77},  075133  (2008).
	
	\bibitem{Wang94}
	X.~Y. Wang, D.~M. Riffe, Y.-S. Lee, and M.~C. Downer, Time-Resolved
	Electron-Temperature Measurement in a Highly Excited Gold Target Using
	Femtosecond Thermoionic Emission, Phys. Rev. B {\bf 50},  8016  (1994).
	
	\bibitem{PeIn13}
	Y.~V. Petrov, N.~A. Inogamov, and K.~P. Migdal, Thermal Conductivity and
	Electron-Ion Heat transfer Coeffcient in Condensed Media with a Strongly
	Excited Electron Subsystem, Pis'ma v Zhurnal Eksperimental'noi i
	Teoreticheskoi Fiziki {\bf 97},  24   (2013).
	
	\bibitem{SkHe13}
	I.~Y. Sklyadneva, R. Heid, K.-P. Bohnen, P.~M. Echenique, and E.~V. Chulkov,
	Mass enhancement parameter in free-standing ultrathin {Pb}(111) films: The
	effect of spin-orbit coupling, Phys. Rev. B {\bf 87},  085440  (2013).
	
	\bibitem{HeLu71}
	L. Hedin and B.~I. Lundqwist, Explicit local exchange-correlation potentials,
	J. Phys. C {\bf 4},  2064  (1971).
	
	\bibitem{Louie79}
	S.~G. Louie, K.-M. Ho, and M.~L. Cohen, Self-consistent mixed-basis approach to
	the electronic structure of solids, Phys. Rev. B {\bf 19},  1774  (1979).
	
	\bibitem{HeBo99}
	R. Heid and K.-P. Bohnen, Linear response in a density-functional mixed-basis
	approach, Phys. Rev. B {\bf 60},  R3709  (1999).
	
	\bibitem{Zein84}
	N.~E. Zein, On density functional calculations of crystal elastic modula and
	phonon spectra, Sov. Phys. Solid State {\bf 26},  1825  (1984).
	
	\bibitem{BaGi01}
	S. Baroni, S. {de Gironcoli}, A. {Dal Corso}, and P. Giannozzi, Phonons and
	related crystal properties from density-functional perturbation theory, Phys.
	Mod. Rev. {\bf 73},  515  (2001).
	
	\bibitem{HeBo10}
	R. Heid, K.~P. Bohnen, I.~Y. Sklyadneva, and E.~V. Chulkov, Effect of
	spin-orbit coupling on the electron-phonon interaction of the superconductors
	{Pb} and {Tl}, Phys. Rev. B {\bf 81},  174527  (2010).
	
	\bibitem{BeSk18}
	G. Benedek, I.~Y. Sklyadneva, E.~V. Chulkov, P.~M. Echenique, R. Heid, K.-P.
	Bohnen, D. Schmicker, S. Schmidt, and J.~P. Toennies, Phonons and
	electron-phonon anomalies in ultra-thin {Pb} films on {Si}(111) and
	{Ge}(111), Surf. Sci. {\bf 678},  38   (2018).
	
	\bibitem{ZaKr17}
	M. Zahedifar and P. Kratzer, Coupling of quantum well states and phonons in
	thin multilayer {Pb} films on {Si(111)}, Phys. Rev. B {\bf 96},  115442
	(2017).
	
	\bibitem{KiBo07}
	P.~S. Kirchmann, M. Wolf, J.~H. Dil, K. Horn, and U. Bovensiepen, Quantum Size
	Effects in {Pb/Si(111)} investigated by laser-induced photoemission, Phys.
	Rev. B {\bf 76},  075406  (2007).
	
	\bibitem{KiBo08}
	P.~S. Kirchmann and U. Bovensiepen, Ultrafast electron dynamics in {Pb/Si(111)}
	investigated by two-photon photoemission, Phys. Rev. B {\bf 78},  035437
	(2008).
	
	\bibitem{KiRe10}
	P.~S. Kirchmann, L. Rettig, X. Zubizarreta, V.~M. Silkin, E.~V. Chulkov, and U.
	Bovensiepen, Quasiparticle lifetimes in metallic quantum-well nanostructures,
	Nature Physics {\bf 6},  782   (2010).
	
	\bibitem{ReKi12}
	L. Rettig, P.~S. Kirchmann, and U. Bovensiepen, Ultrafast dynamics of occupied
	quantum well states in {Pb/Si(111)}, New J. Phys. {\bf 14},  023047  (2012).
	
	\bibitem{ChTr16}
	T. Chase, M. Trigo, A.~H. Reid, R. Li, T. Vecchione, X. Shen, S. Wheatherby, R.
	Coffee, N. Hartmann, A.~D. Reis, X.~J. Wang, and H.~A. D{\"u}rr, Ultrafast
	electron diffraction from non-equilibrium phonons in femtosecond laser heated
	{Au} films, Appl. Phys. Lett. {\bf 108},  041909  (2016).
	
	\bibitem{MaCh19}
	P. Maldonado, T. Chase, A.~H. Reid, X. Shen, R.~K. Li, K. Carva, T. Payer, M.
	{Horn von Hoegen}, K. Sokolowski-Tinten, X.~J. Wang, P.~M. Oppeneer, and
	H.~A. D{\"u}rr, Tracking the Ultrafast Non-Equilibrium Energy Flow between
	Electronic and Lattice Degrees of Freedom in Crystalline Nickel, Phys. Rev. B
	{\bf 101},  100302(R)  (2020).
	
	\bibitem{Seiler2021}
	H. Seiler, D. Zahn, M. Zacharias, P.-N. Hildebrandt, T. Vasileiadis, Y.~W.
	Windsor, Y. Qi, C. Carbogno, C. Draxl, R. Ernstorfer, and F. Caruso,
	Accessing the Anisotropic Nonthermal Phonon Populations in Black Phosphorus,
	Nano Letters {\bf 21},  6171   (2021).
	
\end{thebibliography}

\end{document}